\documentclass[aps,prb,twocolumn,showkeys,groupedaddress]{revtex4}
\usepackage{graphicx}
\usepackage{amsmath,amssymb,amsfonts}
\usepackage{natbib}
\usepackage{color}

\newcommand{\be}{\begin{equation}}
\newcommand{\ba}{\begin{eqnarray}}
\newcommand{\ee}{\end{equation}}
\newcommand{\ea}{\end{eqnarray}}

\begin{document}

\title{Effective Markovian description of decoherence in bound systems}

\author{A. S. Sanz}
\affiliation{Instituto de F{\'\i}sica Fundamental (IFF--CSIC),
Serrano 123, 28006 - Madrid, Spain}

\date{\today}

\begin{abstract}
Effective descriptions accounting for the evolution of quantum systems
that are acted on by a bath are desirable.
As the number of bath degrees of freedom increases and full
quantum simulations turn out computationally prohibitive, simpler models
become essential to understand and gain an insight into the main
physical mechanisms involved in the system dynamics.
In this regard, vibrational decoherence of an I$_2$ diatomics is
tackled here within the framework of Markovian quantum state diffusion.
The I$_2$ dynamics are analyzed in terms of an effective decoherence
rate, $\Lambda$, and the specific choice of the initial state, in
particular, Gaussian wave packets and two-state superpositions.
It is found that, for Markovian baths, the relevant quantity regarding
decoherence is the product of friction ($\eta$) and temperature ($T$);
there is no distinction between varying one or the other.
It is also observed that decoherence becomes faster as the energy
levels involved in the system state correspond to higher eigenvalues.
This effect is due to a population redistribution during the dynamical
process and an eventual irreversible loss of the initial coherence.
These results have been compared with those available in the literature
from more detailed semiclassical IVR simulations, finding a good
agreement.
\end{abstract}


\keywords{Markovian dynamics; quantum state diffusion; vibrational decoherence; dephasing; stochastic quantum trajectory}

\maketitle


\section{\label{sec1} Introduction}

Consider a quantum system (S) coupled to a quantum bath (B).
As it is commonly done in open quantum system theory,\cite{breuer-bk:2002}
let us also assume that initially the system and bath are uncorrelated.
The wave function of the total system, S+B, can then be expressed in
terms of a general, factorizable product state:
\be
 |\Psi\rangle = |\Psi_{\rm S}\rangle \otimes |\Psi_{\rm B}\rangle .
 \label{qs1}
\ee
As time proceeds, the interaction between system and bath makes this
state to become nonseparable or entangled.
This process can be understood as an effective transfer or flow of
coherence between both subsystems, which results in a gradual quenching
of any interference feature associated with the quantum system S.
This phenomenon is what we usually call decoherence.
Typically, it takes place at time scales shorter than energy relaxation,
i.e., before the system energy starts flowing (dissipating) towards the
bath.
Depending on the nature of the bath and the system--bath coupling
strength, the system coherence loss may go from partial (or even total)
reversibility (recurrences) to total irreversibility (Markovianity).
The latter is usually related to the emergence of the classical
world,\cite{giulini-bk,schlosshauer-bk:2007} because the ``classical''
law of addition of probabilities is recovered.

Real systems are not fully isolated in nature.
Therefore, given the major role played by quantum coherence at
microscopic and mesoscopic scales in different areas of physics,
chemistry, and biology, an appropriate characterization of the
system--bath interaction is very important.
Very detailed and accurate simulations of the corresponding systems are
thus essential to study, analyze, and understand the related phenomena
and effects.
This has motivated the development of highly sophisticated numerical
techniques, including classical and semiclassical treatments whenever
the amount of degrees of freedom involved make computationally
prohibitive full quantum-mechanical treatments.
This, however, does not necessarily mean that we fully understand the
system dynamics.
The complexity of these models often makes them intractable from an
analytical viewpoint, hiding the main mechanisms that come into play.
Hence alternative (but complementary) simpler models based on master
equations that capture the essence of the system behavior are
desirable; they provide us with the insight necessary to better
understand such underlying mechanisms.

To illustrate that assertion, consider the Caldeira--Leggett
model used in Wang et al.\cite{miller:JCP-1:2001} to describe the
quenching of vibrational interference effects of an I$_2$ diatomics
coupled to a bath of harmonic oscillators.
This bath is characterized by a spectral density with ohmic form.
The characteristic frequency chosen for this bath is
$\omega_c = 20$~cm$^{-1}$, while the largest frequency of the bath
modes is $\omega_m = 100$~cm$^{-1}$ (a total of 20 to 40 of these
modes were considered in their simulations).
As mentioned above, full quantum simulations are computationally
prohibitive in these cases, so a way to tackle the issue is by means
of the semiclassical initial value representation (SC-IVR)
method.\cite{miller:JCP:1970,miller:FaradayDiscuss:1998}
Apart from its well-known computational advantages, this technique is
also very convenient and interesting regarding the quantum-to-classical
transition: it provides us with a systematic procedure to add quantum
coherence to classical molecular dynamics simulations and therefore
to distinguish between classical and quantum
coherence.\cite{miller:JPCA:2011,miller:JCP:2012}
The analysis reported in Wang et al.\cite{miller:JCP-1:2001} precisely
pursues this simple idea (for a similar application to the two-slit
experiment, see Gelabert et al.\cite{miller:JCP-2:2001})
As the bath temperature and/or system--bath coupling strength are
increased, the system gradually loses its coherence, exhibiting a
``classical-like'' behavior.
Physically, this manifests as a quenching of interference features;
computationally, one finds a better agreement between the SC-IVR and
bare classical molecular dynamics simulations (linearized IVR).

Based on such results, here we address the following question:
can the behaviors reported in Wang et al.\cite{miller:JCP-1:2001} be
properly described in terms of a simple master equation?
Among the different approaches available in the literature to address
this problem,\cite{breuer-bk:2002} we have considered the quantum
state diffusion approach\cite{diosi:JPA:1988,diosi:PLA:1988,diosi:ChemPhys:2001,gisin:JPA:1992,gisin:JPA:1993,gisin:JModOpt:1993,percival-bk}
due to three reasons: ({\it i}) it preserves the appealing concept of
quantum state or wave function as the main descriptor of the system and
its time evolution, ({\it ii}) it stresses the ``flavor'' of the concept
of dephasing, i.e., decoherence not only comes from the interaction with
a bath but also from the incoherence among a large (statistical) number
of realizations, and ({\it iii}) from a technical (numerical) viewpoint,
it is relatively simple to implement (simpler than dealing with density
matrices in configuration space), and yet it still captures the physics
of the phenomenon, \mbox{namely decoherence.}

With respect to {\it ii}, notice that within this theoretical framework,
the expectation value of an operator $\hat{\mathcal{O}}$ that describes
an observable $O$ arises from an average:
\vspace{-0.25cm}
\be
 \langle \hat{\mathcal{O}} \rangle(t) = \frac{1}{N}
 \sum_{i=1}^N \langle\Psi_i(t)|\hat{\mathcal{O}}|\Psi_i(t)\rangle ,
\ee
with the subscript $i$ labeling each particular realization of
the state vector, $|\Psi_i\rangle$, and where $N$ is the total number
of realizations considered; each single realization
$\langle \hat{\mathcal{O}}\rangle_i(t) \equiv \langle\Psi_i(t)|\hat{\mathcal{O}}|\Psi_i(t)\rangle$
describes the time evolution of the expectation value of $\mathcal{O}$
associated with the system state vector $|\Psi_i\rangle$.
Each one of these single realizations constitutes a {\it stochastic
quantum trajectory} analogous to those proposed by Carmichael for
optical systems,\cite{carmichael-bk} although $|\Psi_i\rangle$ displays
a stochastic dynamics throughout its full evolution.
These trajectories describe different physical quantities depending on
which operator $\hat{\mathcal{O}}$ is chosen, being unique for each
$|\Psi_i\rangle$ vector.
This makes an important difference with respect to other types of
quantum trajectories, such as Bohmian trajectories,\cite{sanz-bk-1,sanz-bk-2}
which are not related to a particular
operator, but to a single state vector in configuration representation.

This work has been organized as follows.
The main theoretical aspects of the quantum state diffusion approach
as well as its adaptation to the problem dealt with here are briefly
summarized in Section~\ref{sec2}.
The system and numerical details involved in the simulations considered
here are presented in Section~\ref{sec3}.
In Section~\ref{sec4} the main results are discussed.
Finally, in Section~\ref{sec5}, the main conclusions arising from this
work are summarized.


\section{\label{sec2} Theory}


\subsection{\label{sec21} General aspects}

Consider the dissipative dynamics of a system described within
the framework of the Lindblad approach.\cite{breuer-bk:2002}
Compared with other approaches available in the literature, Lindblad's
one gathers two interesting properties:
\begin{enumerate}
 \item[({\rm i})] it does not arise from any particular representation, thus
 being the most abstract approach that we can find (Redfield's
 approach, for example, comes from the energy representation, although
 later on one can recast it in other representations) and

 \item[({\rm ii})] it provides us with the most general form of equation of motion
 for the density matrix, which satisfies complete positivity.
\end{enumerate}
In this approach, the evolution of the system (reduced) density matrix
$\hat{\rho}_{\rm S}$ is described by the usual Liouvillian term plus a
dissipator, which accounts for the bath effective action over the
system.
More specifically, Lindblad's equation reads as
\begin{equation}
 \frac{d \hat{\rho}_{\rm S}(t)}{dt} =
  - \frac{i}{\hbar} [ \hat{H}, \hat{\rho}_{\rm S}(t) ]
  + \mathcal{D}[\hat{\rho}_{\rm S}(t)] ,
 \label{firstform2}
\end{equation}
where $\hat{H}$ is the Hamiltonian associated with the bare (free)
system and
\begin{equation}
 \mathcal{D}(\hat{\rho}_{\rm S}) \equiv
  \sum_j \Lambda_j \left( 2 \hat{L}_j \hat{\rho}_{\rm S} \hat{L}_j^\dagger
   - \hat{L}_j^\dagger \hat{L}_j \hat{\rho}_{\rm S}
   - \hat{\rho}_{\rm S} \hat{L}_j^\dagger \hat{L}_j \right)
 \label{dissip}
\end{equation}
is the dissipator.
In this latter expression, the $\hat{L}_j$ represent the so-called
Lindblad operators, which describe how the bath specifically acts
over the system.
The relevance of this action is given by the corresponding system--bath
coupling strength constants $\Lambda_j$: the larger the value of
$\Lambda_j$, the faster the decoherence/dissipation induced by
$\hat{L}_j$ occurs.

The Lindblad eq.~\ref{firstform2} can be recast in the form of a
state vector equation\cite{diosi:JPA:1988,diosi:PLA:1988,diosi:ChemPhys:2001,gisin:JPA:1992,gisin:JPA:1993,gisin:JModOpt:1993,percival-bk}
in the It\^o form\cite{ito-bk,ikeda-bk}
\begin{equation}
 |d\Psi\rangle = |v\rangle dt + \sum_j |u_j\rangle d\xi_j .
 \label{itostoch}
\end{equation}
In this diffusion-like equation, $|v\rangle$ is a deterministic
drift term, while the $d\xi_j$ elements denote independent complex
Wiener processes associated with the action of stochastic bath
fluctuations over the system, such that $\langle\Psi|u_j\rangle =0$,
for all $j$, to ensure the state vector normalization.
The means of these Wiener processes over both the distribution and
the fluctuations are assumed to satisfy
\begin{subequations}
 \ba
 E( d\xi_j) & = & 0 , \label{prop1} \\
 E( d\xi_j d\xi_k ) & = & 0 ,
 \label{prop2} \\
 E( d\xi_j^* d\xi_k ) & = & 2 \delta_{jk} dt , \label{prop3}
 \ea
 \label{fluct}
\end{subequations}
i.e., these Wiener processes have zero mean and variance $\sqrt{dt}$.

Physically, eq.~\ref{itostoch} describes a single, stochastic
realization (or time propagation) of the quantum system state vector.
Any expectation value obtained from this realization is called a
{\it stochastic quantum trajectory} because one can keep track of the
corresponding property in time, in analogy to classical
trajectories.\cite{footnote}
In order to obtain an appropriate description of the system dissipative
dynamics, it is necessary to carry out a large number of such
realizations; information about the influence of the bath over the
system (decoherence or dissipation) is extracted from the average
over the corresponding quantum trajectories.
Hence, the bath effects over the system can be somewhat
understood in terms of the dephasing displayed by the different
realizations of the vector state, which leads to its loss of coherence
and eventually also to its relaxation (dissipation).
This picture summarizes the role of the reduced density matrix as
a statistical descriptor of the system dynamics, since it can be
recast~as
\begin{equation}
 \hat{\rho}_{\rm S} = E( |\Psi\rangle \langle \Psi | )
  = \frac{1}{N} \sum_{i=1}^N |\Psi_i\rangle \langle \Psi_i| .
 \label{many}
\end{equation}

Taking into account these facts, particularly eq.~\ref{many},
after some algebra one obtains (see, for example, Gisin and
Percival\cite{gisin:JPA:1992} for a simple derivation) an explicit
functional form for the drift and stochastic terms of
eq.~\ref{itostoch}:
\begin{eqnarray}
 |v\rangle & = & - \frac{i}{\hbar} \ \! \hat{H} | \Psi \rangle
 \nonumber \\ & &
  + \sum_j \Lambda_j \Big( 2 \langle \hat{L}_j^\dagger \rangle_\Psi \hat{L}_j
    - \hat{L}_j^\dagger \hat{L}_j
    - \langle \hat{L}_j^\dagger \rangle_\Psi \langle \hat{L}_j \rangle_\Psi
     \Big) | \Psi \rangle , \nonumber \\ & &
 \label{drift} \\
 |u_j\rangle & = &
 \sqrt{\Lambda_j} \left( \hat{L}_j - \langle \hat{L}_j \rangle_\Psi \right)
   | \Psi \rangle ,
 \label{stoch}
\end{eqnarray}
respectively, with
$\langle \hat{L}_k\rangle_\Psi \equiv \langle\Psi|\hat{L}_j|\Psi\rangle$
being the expectation value of the Lindblad operator $\hat{L}_j$ with
respect to the state vector $|\Psi\rangle$ at time $t$.
Substituting these terms into eq.~\ref{itostoch} gives rise to the
It\^o equation:
\begin{eqnarray}
 |d\Psi\rangle & = & - \frac{i}{\hbar} \ \! \hat{H} | \Psi \rangle dt
 \nonumber \\ & &
  + \sum_j \Lambda_j \Big( 2 \langle \hat{L}_j^\dagger \rangle_\Psi \hat{L}_j
    - \hat{L}_j^\dagger \hat{L}_j
    - \langle \hat{L}_j^\dagger \rangle_\Psi \langle \hat{L}_j \rangle_\Psi
     \Big) | \Psi \rangle dt
 \nonumber \\ & &
  + \sum_j \sqrt{\Lambda_j} \left( \hat{L}_j - \langle \hat{L}_j \rangle_\Psi \right)
   | \Psi \rangle d\xi_j .
 \label{itostoch2}
\end{eqnarray}


\subsection{\label{sec22} Reduced Caldeira--Leggett model}

In order to make practical use of eq.~\ref{itostoch2}, we still
need to define the functional form displayed by the Lindblad operators.
Let us therefore go back to the Caldeira--Leggett model.
Within this model, system and bath are assumed to be coupled
bilinearly, i.e., by means of terms of the form $x X_i$, where $x$
and $X_i$ denote the system and $i$th bath coordinates, respectively.
In the high--temperature regime, this model can be recast in terms
of a master equation:\cite{breuer-bk:2002,accardi-bk}
\begin{equation}
 \frac{d \hat{\rho}_{\rm S}}{dt} =
  - \frac{i}{\hbar} \ \! [ \hat{H},\hat{\rho}_{\rm S} ]
  - \frac{i\eta}{\hbar} \ \!
    [ \hat{x}, \{ \hat{p},\hat{\rho}_{\rm S} \} ]
  - \frac{2m\eta k_B T}{\hbar^2} \ \!
    [ \hat{x}, [ \hat{x},\hat{\rho}_{\rm S} ] ]
 \label{cl}
\end{equation}
where the first term represents the system unitary evolution, the
second its relaxation, and the third its decoherence.
Although this equation is Markovian, it is not of the Lindblad form.
Nonetheless, at high temperatures, one can add to (\ref{cl}) the term
\begin{equation}
 - \frac{\eta}{8m k_B T} \ \! [ \hat{p}, [ \hat{p},\hat{\rho}_{\rm S} ] ] ,
 \label{additional}
\end{equation}
which is relatively small.
The Lindblad form appears after diagonalizing the corresponding
dissipator, which renders the associated Lindblad operators.\cite{breuer-bk:2002}

A simpler way to proceed than diagonalizing, however, consists of
assuming that at relatively classical regimes, i.e., when Planck's
constant is relatively small compared with the actions involved and the
object is massive, the third term on the right-hand side of
eq.~\ref{cl} is much larger than the second one.
This allows us to simplify eq.~\ref{cl}\cite{zurek:PRL:1993,joos:bk:1996}~as
\begin{eqnarray}
 \frac{d \hat{\rho}_{\rm S}}{dt} & = &
  - \frac{i}{\hbar} \ \! [ \hat{H},\hat{\rho}_{\rm S} ]
  - \frac{2m\eta k_B T}{\hbar^2} \ \!
    [ \hat{x}, [ \hat{x},\hat{\rho}_{\rm S} ] ]
  \nonumber \\
 & = & - \frac{i}{\hbar} \ \! [ \hat{H},\hat{\rho}_{\rm S} ]
  + \frac{2m\eta k_B T}{\hbar^2} \ \!
    \Big( 2 \hat{x} \hat{\rho}_{\rm S} \hat{x}
     - \hat{x}^2 \hat{\rho}_{\rm S}
     - \hat{\rho}_{\rm S} \hat{x}^2 \Big) .
  \nonumber \\ & &
 \label{cl3}
\end{eqnarray}
This equation, already in the Lindblad form, is known as the
reduced Caldeira--Leggett model.\cite{breuer-bk:2002}
Its range of validity can be easily inferred as follows.
The hypothesis considered to obtain (\ref{cl3}) is equivalent to the
previous assumption that (\ref{additional}) is small compared with the
third term of (\ref{cl}).
On the other hand, if $\omega$ is some characteristic frequency
associated with the system, one would expect that the corresponding
momenta go like $p \sim m\omega x$, approximately.
Substituting this estimate of the momentum into eq.~\ref{additional}
and then making use of the above assumptions, we find
\be
 \frac{4k_BT}{\hbar\omega} \gg 1 .
 \label{validity}
\ee
Accordingly, eq.~\ref{cl3} is valid whenever temperatures satisfy
this relation.

Following the state vector approach introduced in Section~\ref{sec21},
eq.~\ref{cl3} can be recast as a quantum stochastic differential
equation, namely:
\ba
 |d\Psi\rangle & = & - \frac{i}{\hbar} \ \! \hat{H} |\Psi\rangle dt
 - \Lambda \left( x - \langle x \rangle \right)^2 |\Psi\rangle dt
 \nonumber \\ & &
 + \sqrt{\Lambda} \left( x - \langle x \rangle \right)
  |\Psi\rangle d\xi ,
 \label{qsde}
\ea
with one Lindblad operator, $\hat{L} = \sqrt{\Lambda}\ \! \hat{x}$,
and where the system--bath coupling strength is given by the decoherence
rate:
\be
 \Lambda = \frac{2m\eta k_B T}{\hbar^2} ,
 \label{lrate}
\ee
with units of (space)$^{-2}\times$(time)$^{-1}$.
This rate provides us with an estimate of the time scales at which
the correlation (coherence) between two points in configuration
space is lost.
For example, for two points separated a distance $\ell$, this
time scale will be of the order of $1/\Lambda\ell^2$.

The numerical simulations reported in Section~\ref{sec4} constitute
a test of the feasibility and applicability of eq.~\ref{cl3} to
study in a simplified fashion systems afforded by the full
Caldeira--Leggett model.
Notice that in the ranges of temperature where this description is
valid, one has an interesting, effective tool to probe open quantum
system dynamics, where all bath effects (temperature and friction) are
enclosed within a single parameter, namely the decoherence rate
$\Lambda$.
Because the bath dynamics are not explicitly considered, there is a
remarkable reduction of the computational time demand with respect
to full system--bath treatments (classical or semiclassical).
In this sense, it is worth stressing that the state vector approach
could be advantageously used to explore the system dynamics in some
ranges of parameters of interest, previous to full, more detailed
dynamical simulations.
On the contrary, as a feedback, the latter type of calculations could
be used to design and implement better Lindblad operators that would
help to improve the model based on eq.~\ref{cl3} and eventual
interpretations relying on it.


\section{Numerical details}
\label{sec3}

As in Wang et al.\cite{miller:JCP-1:2001}, here we have analyzed the
gradual coherence loss displayed by the radial distribution function of
the I$_2$.
The lowest electronic energy surface describing this system can be
modeled by a Morse function along the radial direction (here
denoted by $x$):
\begin{equation}
 V(x) = D \left[ 1 - e^{- \alpha (x - x_e)} \right]^2 ,
 \label{eq22}
\end{equation}
with parameters $D = 1.2547\times10^4$~cm$^{-1}$,
$\alpha = 1.8576$~\AA$^{-1}$, and $x_e = 2.6663$~\AA.
This Morse oscillator supports about 120 bound states and has
a harmonic frequency
\begin{equation}
 \omega_0 = \sqrt{\frac{2\alpha^2D}{m}} \approx 214.6~{\rm cm}^{-1}
  = 40.451~{\rm ps}^{-1} ,
 \label{harmfreq}
\end{equation}
where $m$ is the I$_2$ reduced mass ($m = m_0/4$, with $m_0=4.22 \times
10^{-22}$~g being the I$_2$ mass).
To compare with Wang et al.\cite{miller:JCP-1:2001}, first we have considered the
dynamics displayed by a Gaussian wave packet:
\begin{equation}
 \Psi_0 (x) = \left( \frac{1}{2\pi\sigma^2} \right)^{1/4}
  e^{-(x-x_0)^2/4\sigma_0^2 + ip_0(x-x_0)/\hbar} ,
 \label{wave0}
\end{equation}
with parameters $x_0 = 2.4$~\AA, $p_0 = 0$, and $\sigma_0^2 =
\hbar/2m\omega_0$, under the action of the Morse potential (eq.~\ref{eq22})
and a stochastic noise satisfying the properties of eq.~\ref{fluct}.
Several two-state superpositions have also been studied in order to
understand the relationship between coherence and population dynamics.

The scheme followed to solve numerically eq.~\ref{qsde} consists of
attacking separately the Hamiltonian and the diffusive parts and then
combining them together,\cite{brumer-gong:PRE:1999,prezhdo:PRL:2000,brumer-han:JCP:2005}
thus following a strategy somewhat analogous to operate in the
interaction picture.
The Hamiltonian part (first term on the right-hand side of
eq.~\ref{qsde}) is integrated by making use of the split-operator
scheme\cite{feit-fleck:JCompPhys:1982,feit-fleck:JCP:1983,feit-fleck:JCP:1984}
combined with the fast Fourier method.\cite{NumRecipes-bk}
The diffusive part [second and third terms on the right-hand
side of eq.~\ref{qsde}] is separately integrated with a
second-order Runge-Kutta algorithm adapted to stochastic processes.\cite{klauder:SIAM:1985}
The updated wave function results from the addition of both solutions.
A single realization of the state vector $|\Psi\rangle$ is
obtained by proceeding recursively in this manner until concluding
the time propagation.
In the calculations, a total of 2500 realizations for the Gaussian wave
packet and 2000 for the superpositions have been considered.
These numbers have been found to be optimal for the quantities
computed and presented here, although even more realizations could be
necessary in other cases.
This happens, for example, in the calculation of energy-level
populations and coherences for the Gaussian wave packet (\ref{wave0}).
The energy levels involved in this Gaussian state are higher
than those intervening in the two-state superpositions considered below,
so obtaining smoothly converged quantum trajectories for them requires
a higher number of realizations than for the latter.


\section{Results and discussion}
\label{sec4}


\subsection{Wave packet dynamics}
\label{sec41}

\begin{figure}[t]
 \includegraphics[width=7cm]{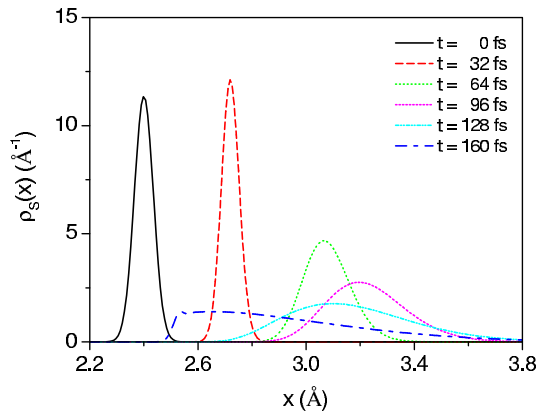}
 \caption{\label{fig1-CJC}
  Snapshots of the (averaged) reduced probability density within
  the first vibrational period (see Fig.~\ref{fig2-CJC}a) inside the
  Morse potential (eq.~\ref{eq22}) for negligible decoherence
  ($\Lambda = 10^{-8}$~a.u.).
  The initial state corresponds to the Gaussian wave packet
  (eq.~\ref{wave0}).
  The times at which each snapshot was taken are indicated in
  the legend within the figure.}
\end{figure}

\begin{figure}[t]
 \includegraphics[width=7cm]{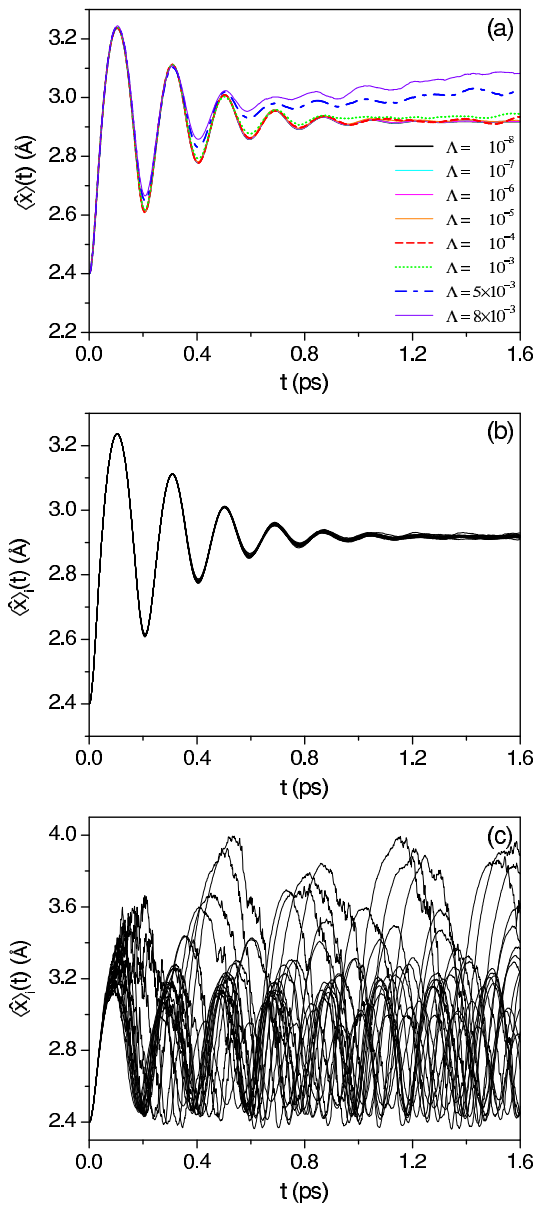}
 \caption{\label{fig2-CJC}
  (a) Position expectation value for different decoherence rates,
  from $\Lambda = 10^{-8}$ to $8\times10^{-3}$~a.u.
  (b) Set of 20 quantum trajectories, $\langle \hat{x}\rangle_i(t)$,
  associated with the first 20 realizations of the Gaussian wave
  packet (eq.~\ref{wave0}) for $\Lambda = 10^{-8}$~a.u.
  (c) The same as in Fig.~\ref{fig2-CJC}b, but for $\Lambda = 5\times10^{-3}$~a.u.}
\end{figure}

In order to test the accuracy and stability of the numerical
algorithm, first a trial simulation has been run with the initial
wave packet (\ref{wave0}) and $\Lambda = 10^{-8}$ (given for simplicity
in atomic units: 1~a.u.\ ($\Lambda$) = 147.6~\AA$^{-2}$fs$^{-1}$
$\approx 1.5\times 10^{32}$~cm$^{-2}$s$^{-1}$).
This $\Lambda$ value is relatively small to have important effects on
the unitary part of the algorithm and therefore allows us to obtain a
close solution to a noise-free ($\Lambda = 0$) propagation.
A series of snapshots of the averaged probability density
\ba
 \rho_{\rm S}(x,t) & = & \langle x|\hat{\rho}_{\rm S}(t)|x\rangle
 \nonumber \\
 & = & \langle x| \left[ \frac{1}{N} \sum_{i=1}^N
     |\Psi_i(t)\rangle \langle\Psi_i(t)| \right] |x\rangle
 \nonumber \\
 & = & \frac{1}{N} \sum_{i=1}^N \langle x|
     \Psi_i(t)\rangle \langle\Psi_i(t) |x\rangle ,
\ea
with $N=2500$,
spanning a time of 160~fs is displayed in Fig.~\ref{fig1-CJC}.
This time covers the first harmonic vibrational period
($\tau_0 = 2\pi/\omega_0 \approx 155.3$~fs), although it is about three
fourths of the oscillation period for the system considered here
(see Fig.~\ref{fig2-CJC}a).
The discrepancy between these two characteristic vibrational
periods is due to the anharmonicity of the Morse potential function.
Notice that the wave packet energy expectation value
$\langle \hat{H} \rangle$ is about $0.4 D$.
In such cases, the oscillation frequency between the two turning
points of the Morse potential at a certain energy $E$ is given by
\cite{deMarcus:AJ:1978}
\be
 \omega_M = \omega_0 \sqrt{1 - \frac{E}{D}} .
 \label{freqanh}
\ee
Assuming that $E \sim \langle \hat{H} \rangle$, we obtain
$\omega_M \sim 0.77 \omega_0$, in agreement with the previous
statement.
The dynamics are therefore quite anharmonic, as seen in the
figure: the wave packet spreading increases significantly along
the propagation, contrary to the frozen oscillatory behavior displayed
by the same wave packet in a harmonic potential.\cite{sanz:cpl:2007}
This implies that, after some time, the foremost part of the wave
packet will bounce backwards and overlap with the rearmost one, giving
rise to the emergence of interference features (see the wave packet
denoted with a blue dashed-dotted line).

The position expectation value $\langle \hat{x}\rangle(t)$ is
represented in Fig.~\ref{fig2-CJC}a and its oscillatory behavior
provides us with a characteristic time scale for the system, namely
$\bar{\tau} \approx 200$~fs, which is in agreement with the value
obtained through eq.~\ref{freqanh}, i.e., $\tau_M \sim 1.3\tau_0$.
After substitution of the associated frequency
$\bar{\omega} = 2\pi/\bar{\tau}$ into eq.~\ref{validity}, we find
that the validity of the state vector approach is ensured in our
case for $T \gg 60$~K.
The curves displayed in this figure also provide us with an idea
of how $\Lambda$ influences the wave packet dynamics.
As can be seen, for about two periods the position expectation
value is not very sensitive to $\Lambda$; the damping observed in
the oscillations is much more related to the anharmonicity of the
potential than to decoherence.
Decoherence effects become more relevant at around
$t \approx 400-500$~fs, particularly for $\Lambda > 10^{-3}$.
Notice that, in the long term, $\langle \hat{x}\rangle(t)$ does not
approach the center of the well but displays a slight deviation
to its right --- towards the ``softer'' part of the Morse well.
These values of $\Lambda$ are of the order of
10$^{28}$~cm$^{-2}$s$^{-1}$ or higher, in agreement with the
estimates provided by Joos and Zeh\cite{joos-zeh:ZPhysB:1985} for
large molecules (with a size of $\sim 10^{-6}$~cm) acted on by air
molecules at $T=300$~K: $\Lambda \sim 10^{30}$~cm$^{-2}$s$^{-1}$.

The explanation for such a behavior can be readily understood by
inspecting Figs.~\ref{fig2-CJC}b and \ref{fig2-CJC}c for
$\Lambda = 10^{-8}$ and $\Lambda = 5\times 10^{-3}$, respectively.
In each panel, a set of 20 quantum trajectories, $\langle \hat{x}\rangle_i(t)$,
is displayed.
For $\Lambda = 10^{-8}$, these trajectories essentially behave in the
same way, not showing relevant deviations when one is compared with the
others.
In contrast, for $\Lambda = 5\times 10^{-3}$, there are trajectories
that display larger excursions towards the softer region of the
potential well.
It is this behavior that eventually leads to the outwards
displacement of the asymptotic value of $\langle \hat{x} \rangle$
observed in Fig.~\ref{fig2-CJC}a.

\begin{figure}[t]
 \includegraphics[width=7cm]{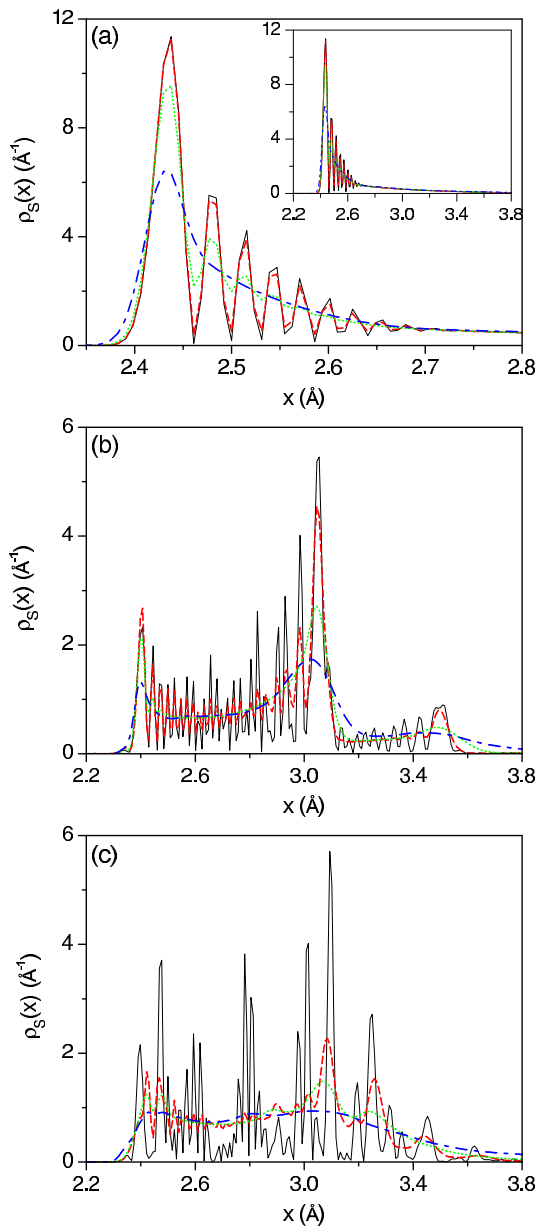}
 \caption{\label{fig3-CJC}
  Reduced probability density for different decoherence rates:
  $\Lambda = 10^{-8}$~a.u.\ (black solid line),
  $\Lambda = 10^{-4}$~a.u.\ (red dashed line),
  $\Lambda = 10^{-3}$~a.u.\ (green dotted line), and
  $\Lambda = 5\times10^{-3}$~a.u.\ (blue dashed-dotted line).
  (a) $t = 192$~fs, (b) $t = 640$~fs, and (c) $t = 1600$~fs.
  In Fig.~\ref{fig3-CJC}a, the inset shows the long reach of the tail
  of the wave packet at $t=192$~fs.}
\end{figure}

To study now the decoherence effects induced by the bath on the
system, three snapshots of the probability density have been
represented in Fig.~\ref{fig3-CJC}.
The interference oscillations that can be seen in the three
panels for $\Lambda=10^{-8}$ constitute a signature of the system
spatial quantum coherence, which is gradually lost as the system--bath
coupling strength, $\Lambda$, increases.
Thus, for $\Lambda = 10^{-3}$ and $\Lambda = 5\times 10^{-3}$,
a seemingly classical behavior is apparent: interference oscillations
are totally suppressed and the distribution seems to be an ``average''
over the mean value of the totally coherent probability density.
Indeed, it is this behavior that makes the right-hand-side ``tail'' of
the distribution extend farther away for long times, provoking
the deviation of $\langle \hat{x}\rangle(t)$ beyond 2.9~\AA\
(see Fig.~\ref{fig2-CJC}a).

Comparing Fig.~\ref{fig3-CJC} with figs.~4 and~5 from Wang et
al.\cite{miller:JCP-1:2001}, we find that the state vector
model is in good agreement with the forward-backward (FB) IVR results,
which explicitly include the quantum dynamics of the bath degrees of
freedom (described as Gaussian wave packets).
In the simulations performed in Wang et al.\cite{miller:JCP-1:2001},
decoherence is analyzed essentially in terms of two bath-related
parameters, namely friction ($\eta$) and temperature ($T$).
Friction affects the system--bath coupling strength through the
coupling coefficients $c_j$ in the full Hamiltonian; temperature
is varied by selecting the initial conditions according to a Boltzmann
distribution at the corresponding temperature.
In our case, as mentioned above, decoherence only depends on the
rate $\Lambda$, since both $\eta$ and $T$ are included in it, as seen
in (\ref{lrate}).
Nevertheless, even though we cannot make a separate analysis (which
would allow us to determine the negligible effects of a finite coupling
at $T=0$, for example), a reliable comparison with the FB-IVR results
is still possible by considering the quantity
$\xi = \eta_e T$ (with $\eta_e \equiv \eta/m\omega_0$
\cite{miller:JCP-1:2001}).

Taking into account the above facts, we note that the state vector
approach effectively captures and reproduces the physics of the more
detailed FB-IVR analysis.
For example, comparing Fig.~\ref{fig3-CJC}a with fig.~4 from Wang et
al.\cite{miller:JCP-1:2001}, we find that the calculation with $\Lambda =
10^{-3}$~a.u.\ produces an interference quenching similar to that
observed for the cases with $\eta_e=0.15$ and $T=100$~K and
$\eta_e=0.05$ and $T=300$~K for which $\xi = 15$.
This result supports the fact that, at least in the case of thermal
baths describable in terms of an ohmic spectral density, the crux of
the matter regarding decoherence is the product $\eta T$ (or,
equivalently, $\eta_e T$), as already pointed out by Elran and
Brumer\cite{brumer-elran:JCP:2004} by also using the FB-IVR method.
That is, no matter which one of the two parameters is varied, the
same decoherence effects will be essentially observed, provided their
product $\xi$ remains constant.
This fact reinforces the use of the quantum state diffusion model,
particularly eq.~\ref{qsde}, where $\eta$ and $T$ appear combined
together within the decoherence rate $\Lambda$.
This property can be used to find out other analogous behaviors.
For example, consider the cases for $\eta_e=0.25$ with $T=100$ and
300~K.
For these, we find $\xi_{100}=15$ and $\xi_{300}=75$, respectively,
with their ratio being $\xi_{300}/\xi_{100} = 5$.
Given the proportionality between $\xi$ and $\Lambda$, if
$\xi=15$ approximately corresponds to $\Lambda = 10^{-3}$,
then $\xi=75$ will correspond to a decoherence rate five times
larger, i.e., $\Lambda = 5\times 10^{-3}$.
This is, effectively, what can be observed when the green dashed-dotted
line in Fig.~\ref{fig3-CJC}a is compared with the case with
$\eta_e=0.25$ and $T=300$~K displayed in fig.~4 of Wang et
al.\cite{miller:JCP-1:2001}

For longer time scales (see Figs.~\ref{fig3-CJC}b and \ref{fig3-CJC}c),
we already start noticing a smearing out of the probability density,
which is not present in the FB-IVR results, although it is consistent
with the typical exponential-like decay undergone by quantum coherence
according to eq.~\ref{cl3} \cite{joos:bk:1996}.
Let us consider the following definition of coherence
length:\cite{joos-zeh:ZPhysB:1985}
\be
 \ell(t) \equiv \left(\sqrt{8\Lambda t}\right)^{-1/2} ,
 \label{cohl}
\ee
which provides us with an estimate of the distance along which
coherence is still preserved, and therefore information about the
quenching of interference features.
In Table~\ref{tab1} some estimates of the coherence length are
given for $\Lambda = 10^{-4}$ and $10^{-3}$ at different times.
In the case of $\Lambda = 10^{-4}$, as seen in the three panels of
Fig.~\ref{fig3-CJC}, the corresponding coherence lengths cover the
width of at least several interference oscillations at their
respective times.
Hence interference features are still apparent even at $t=1600$~fs.
For $\Lambda = 10^{-3}$, however, only at $t=192$~fs we can observe a
series of weak interference oscillations, since the spatial coherence
hardly covers the width of one oscillation.
At later times, very weak interference features can be seen around
$x \sim 2.4$~\AA, where the width of the oscillations is still
comparable with the coherence length.

Now, why is there a difference with respect to the FB-IVR simulations
in spite of the consistency shown by the model?
Here, one could be tempted to think whether the use of a limited set
of harmonic oscillators in the FB-IVR is not producing a ``fake''
recoherence effect, since the Caldeira--Leggett model requires, in
principle, an infinite collection of them.
In other words, the description with a few oscillators may be valid
for short times, but not for longer ones, as one may infer from
the classical Wigner method.\cite{brumer-elran:JCP:2013}
For example, the largest frequency associated with these modes
($\omega_m = 100$~cm$^{-1}$) involves a characteristic time scale
about twice as large as the Morse harmonic period.
Therefore, after a number of such periods, it is reasonable to expect
the appearance of recurrences, which may play a role by putting some
coherence back into the system.
Notice that the interaction between the system and each bath particle
is relatively simple, and that there are no intrabath couplings,
which at high temperatures give rise to faster decoherence rates.\cite{sanz:PRE:2012}
This is easy to understand.
As the bath temperature increases, not only is a faster
transfer of coherence from the system to the bath expected but also
that this coherence is more effectively transferred among different
bath particles, something that cannot happen in the Caldeira--Leggett
model because of its lack of intrabath couplings.
In this sense, although a small number of bath oscillators (particles)
seems to suffice for convergence in semiclassical simulations of
the Caldeira--Leggett model,\cite{miller:JCP-1:2001,brumer-elran:JCP:2004,grossmann:JCP:2009}
avoiding possible classical recurrences, some coherence effects may
still be active due to a lower transfer rate of this coherence among
the bath degrees of freedom.
Nevertheless, as acknowledged by Elran and Brumer,\cite{brumer-elran:JCP:2004}
it is also important to keep in mind
that because of the hypothesis upon which Markovian models are built,
they cannot cope with memory or backreaction.
These effects may play an important role in the system dynamics, for
example, at low temperatures or for a small number of bath particles
(and relatively weak system--bath couplings).
In this sense, there must always be a balance between one kind of
approach and the others (full dynamical models and effective ones).
They constitute different alternatives, but must also be complementary.

\begin{table}[t]
 \caption{\label{tab1} Coherence length (eq.~\ref{cohl}) for different
  values of the decoherence rate and time.}
 \begin{tabular}{c p{.3cm} c p{.3cm} c}
  \hline\hline
   $\Lambda$ (a.u.) & & $t$ (fs)  & & $\ell$ (\AA) \\ \hline
                    & &  \phantom{0}192   & &    0.21\phantom{0} \\
       10$^{-4}$    & &  \phantom{0}640   & &    0.11\phantom{0} \\
                    & &   1600            & &    0.073     \\ \hline
                    & &  \phantom{0}192   & &    0.066     \\
       10$^{-3}$    & &  \phantom{0}640   & &    0.036     \\
                    & &   1600            & &    0.023     \\ \hline\hline
 \end{tabular}
\end{table}

\begin{figure*}[t]
 \includegraphics[width=17cm]{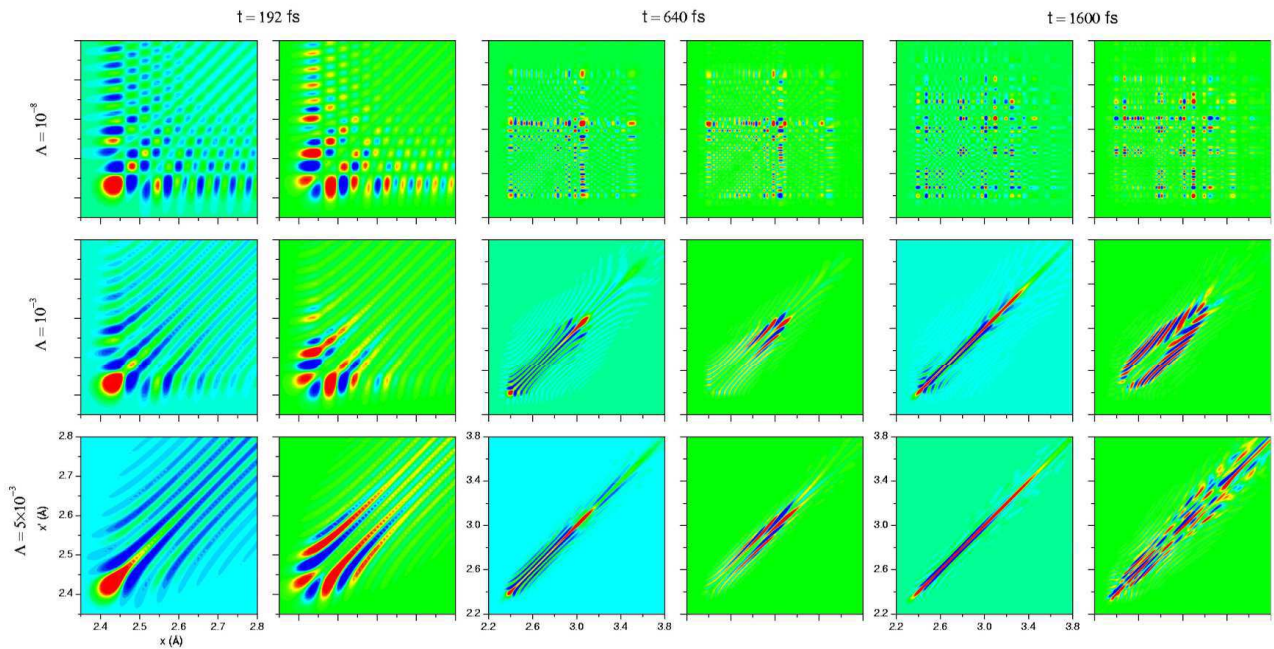}
 \caption{\label{fig4-CJC}
  Density matrix in configuration space at $t=192$, 640, and 1600~fs.
  At each time, left/right panels represent the real/imaginary part of
  the density matrix.
  Different decoherence rates: $\Lambda = 10^{-8}$, $10^{-3}$, and
  $5\times10^{-3}$.
  The color scale, from blue to red, indicates the transition from
  minimum (negative) to maximum (positive) values.}
\end{figure*}

A plot of the reduced density matrix in configuration space,
$\rho_{\rm S}(x,x',t)$, also provides us with another interesting
perspective on the decoherence process.
In Fig.~\ref{fig4-CJC}, the real and imaginary parts of this matrix
(left and right columns, respectively, for each time) have been plotted
at the same three times considered in Fig.~\ref{fig3-CJC} and for three
decoherence regimes: negligible ($\Lambda = 10^{-8}$), moderate
($\Lambda = 10^{-3}$), and strong ($\Lambda = 5\times 10^{-3}$).
The real part of the reduced density matrix displays even symmetry
with respect to the diagonal axis, $x=x'$, while its imaginary
part is odd;
the diagonal of the real part corresponds to the probability
density (see Fig.~\ref{fig3-CJC}a), while the diagonal of the
imaginary one vanishes.
In both cases, the most remarkable feature is the chessboard-like
structure due to interference.
As $\Lambda$ increases, this structure fades out and a stripe-like
structure emerges.
These stripes, parallel to the axis $x=x'$, denote the persistence of
some amount of quantumness.
As the value of $\Lambda$ increases even more, this structure also
disappears, only surviving the terms close to the diagonal (even
though some small off-diagonal contributions can still persist in
the imaginary part of the reduced density matrix).


\subsection{Two-state superposition dynamics}
\label{sec42}

The analysis of the system dynamics studied in the previous section
in terms of populations and coherences is also very interesting,
since it takes us from the configuration space to an energy
representation.
When the Gaussian wave packet of the previous section is recast
as a superposition of eigenfunctions of the Morse
oscillator,\cite{morse:PhysRev:1929,muendel:ChemPhys:1986}
about 70 of the approximately 120 bound states supported by this
potential contribute to the wave packet dynamics.
Analyzing the behavior of the associated populations and coherences
will then be more confusing than clarifying.
Hence, instead, we are going to consider a series of different two-state
superpositions, in particular with low-energy states, for which a total
of 2000 realizations is enough to obtain converged results.
As the initial wave function, we will use
\begin{equation}
 \Psi_{mn,0} (x) = c_m \Phi_m (x) + c_n \Phi_n (x) ,
 \label{wave02}
\end{equation}
where $m$ and $n$ label the corresponding Morse eigenfunctions.
In all cases, we have considered ({\rm i}) $n>m$,
({\rm ii}) $c_m^2 = 0.4$
and $c_n^2 = 0.6$ to have a biased superposition,
and ({\rm iii}) a moderate decoherence regime, with
$\Lambda = 9\times 10^{-3}$~\AA$^{-2}$fs$^{-1}$
($\sim 6.1\times 10^{-5}$~a.u.), which produces a substantial quenching
of the interference features along the propagation, but without fully
suppressing them ($\ell \approx 0.068$~\AA).

\begin{figure}[!t]
 \includegraphics[width=7cm]{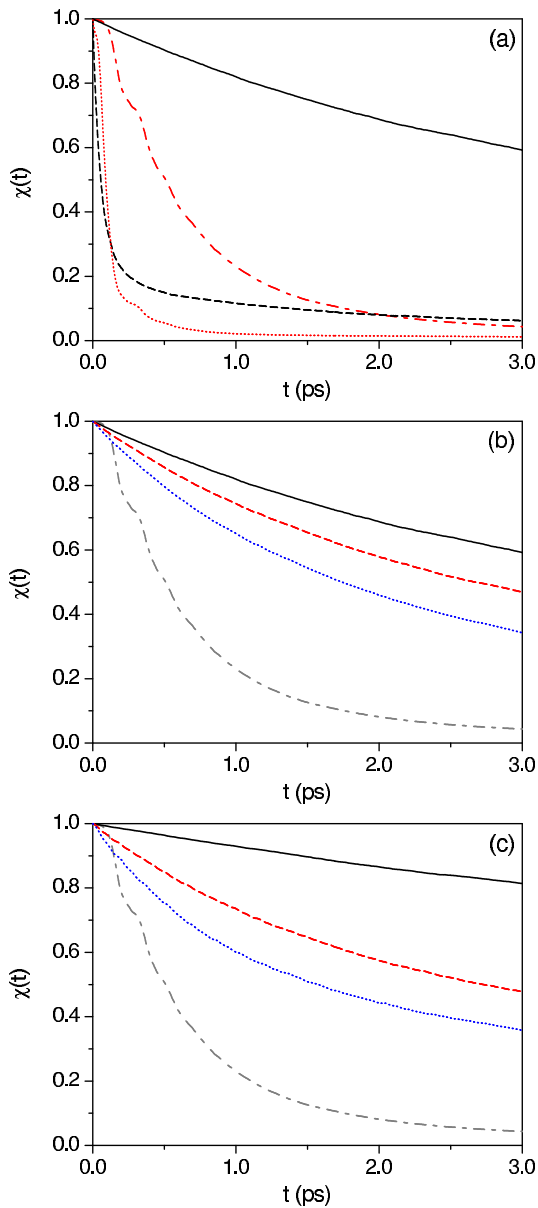}
 \caption{\label{fig5-CJC}
  (a) Time evolution of $\chi$ for $\Psi_{03}$ and
  $\Lambda = 9\times 10^{-3}$~\AA$^{-2}$fs$^{-1}$ (black solid line),
  $\Psi_{03}$ and $\Lambda \approx 0.74$~\AA$^{-2}$fs$^{-1}$
  (black dashed line), a Gaussian and
  $\Lambda = 9\times 10^{-3}$~\AA$^{-2}$fs$^{-1}$
  (red dash-dotted line), and a Gaussian and
  $\Lambda \approx 0.74$~\AA$^{-2}$fs$^{-1}$ (red dotted line).
  (b) Time evolution of $\chi$ for $\Psi_{03}$ (black solid line),
  $\Psi_{05}$ (red dashed line), and $\Psi_{36}$ (blue dotted line).
  (c) Time evolution of $\chi$ for $\Psi_{01}$ (black solid line),
  $\Psi_{45}$ (red dashed line), and $\Psi_{89}$ (blue dotted line).
  In Figs.~\ref{fig5-CJC}b and \ref{fig5-CJC}c,
  $\Lambda = 9\times 10^{-3}$~\AA$^{-2}$fs$^{-1}$ and, to compare with,
  the time evolution of $\chi$ for a Gaussian wave packet (gray
  dashed-dotted line) under the same conditions has also been included.}
\end{figure}

The first quantity of interest that we are going to analyze is the
purity:
\ba
 \chi(t) & = & {\rm Tr} [\hat{\rho}_{\rm S}^2(t)]
  = \sum_{i,j} \langle \Phi_i|\hat{\rho}_{\rm S}(t)|\Phi_j\rangle
    \langle \Phi_j|\hat{\rho}_{\rm S}(t)|\Phi_i\rangle \nonumber \\
 & = & \sum_i |\langle \Phi_i|\hat{\rho}_{\rm S}(t)|\Phi_i\rangle|^2
      + \sum_{i\ne j} |\langle \Phi_j|\hat{\rho}_{\rm S}(t)|\Phi_i\rangle|^2 ,
 \nonumber \\ & &
 \label{trace}
\ea
which is a measure of the degree of ``mixedness'' of the system
quantum state\cite{schlosshauer-bk:2007} or, within our context, of
the incoherence among different realizations.
Notice that this dephasing as well as the fact that other states
apart from $\Phi_n$ and $\Phi_m$ may become populated with time will
lead to $\chi(t) \le 1$ as time increases
(at $t=0$ and/or $\Lambda=0$, $\chi = c_n^2 + c_m^2 = 1$).
To understand the sensitivity of the decoherent process to $\Lambda$
and more specifically the choice of the initial state, in
Fig.~\ref{fig5-CJC}a, we have plotted the time evolution of $\chi$
for a (0,3)-superposition and the Gaussian wave packet considered in
Section~\ref{sec41} for two values of $\Lambda$.
As can be seen, for a given value of $\Lambda$, the larger amount
of eigenstates involved in the dynamics of the Gaussian wave packet
produces a decay of its purity faster than for $\Psi_{03}$.
Actually, while $\chi$ decays smoothly for the superposition, a series
of steps or oscillations are noticeable in the case of the Gaussian
state.
This decay, as is seen for higher $\Lambda$, takes place in
two time scales rather than one, as confirmed by a best-fit analysis
with one and two decaying exponential functions.
In the first case, decay times of about 111~fs for $\Psi_{03}$ and
103~fs for the Gaussian have been obtained, with fitting correlation
factors of 0.928 and 0.979, respectively.
In the second case, decay times of about 61 and 998~fs for $\Psi_{03}$,
and 94 and 804~fs for the Gaussian have been obtained,
with correlations of 0.999 and 0.982, respectively (obviously, the
substantially better agreement for $\Psi_{03}$ was expected due to the
lack of oscillations in its $\chi$-function).
Again, these facts can also be determined from FB-IVR calculations,
as shown by Elran and Brumer\cite{brumer-elran:JCP:2004} (see
fig.~4 in this work, when the fluctuating behavior of the graphs is
neglected and one focuses only on their average trend).

In order to investigate now the bath effects on the system
depending on the components of the initial superposition, in
Fig.~\ref{fig5-CJC}b, we have plotted $\chi(t)$ for $\Psi_{03}$,
$\Psi_{05}$, and $\Psi_{36}$ (for comparison, the graph for the
Gaussian state has also been included).
As can be seen, the decay of the purity becomes faster as the
components forming the superposition are higher in energy.
This explains why the purity for the Gaussian state decays so quickly
with respect to the superpositions: initially, considering only those
states whose populations are $\ge 0.01$, we have a superposition of 34
eigenstates, from $\Phi_{12}$ to $\Phi_{46}$.
Now, given that these eigenstates are consecutive in energy, one may
ask about the decay for two-state superpositions of this kind.
Results for the superposition states $\Psi_{01}$, $\Psi_{45}$, and
$\Psi_{89}$ are displayed in Fig.~\ref{fig5-CJC}c and, as expected,
as the energy increases, the decay of $\chi$ becomes faster.
However, this decay is not homogeneous, although the energy levels
are consecutive in all superpositions.
As can be noticed, the distance between $\chi_{01}$ and $\chi_{45}$
is larger than between $\chi_{45}$ and $\chi_{89}$.
This could be connected to the fact that the relative difference,
defined as
\be
 \Delta_{mn} = \left( \frac{En - E_m}{E_n} \right) \times 100\% ,
\ee
is larger in the case of the (0,1)-superposition ($\sim 66\%$) than for
the other two ($\sim 22\%$ for $\Phi_{45}$ and $\sim 11\%$ for
$\Phi_{89}$).
A smaller difference between energy levels means a larger recurrence
time $\tau_{mn} = 2\pi\hbar/(E_n - E_m)$ and, therefore, a higher
susceptibility to be acted by decoherence, which is precisely what
we observe in the figure (notice that $\tau_{01} \approx 157$~fs,
$\tau_{45} \approx 162$~fs, and $\tau_{89} \approx 167$~fs).
From a best-fit to a single decaying exponential, we find that the
decay times are $\bar{\tau}_{01} \sim 6.4$~ps for $\Phi_{01}$,
$\bar{\tau}_{45} \sim 2.0$~ps for $\Phi_{45}$, and
$\bar{\tau}_{89} \sim 1.2$~ps for $\Phi_{89}$.
The ratios of these characteristic times,
$\bar{\tau}_{01}/\bar{\tau}_{45} \sim 3.2$ and
$\bar{\tau}_{45}/\bar{\tau}_{89} \sim 1.7$, actually seem to be
consistent with the ratios between relative differences,
$\Delta_{01}/\Delta_{45} \sim 3$ and $\Delta_{45}/\Delta_{89} \sim 2$.

\begin{figure}[t]
 \includegraphics[width=7cm]{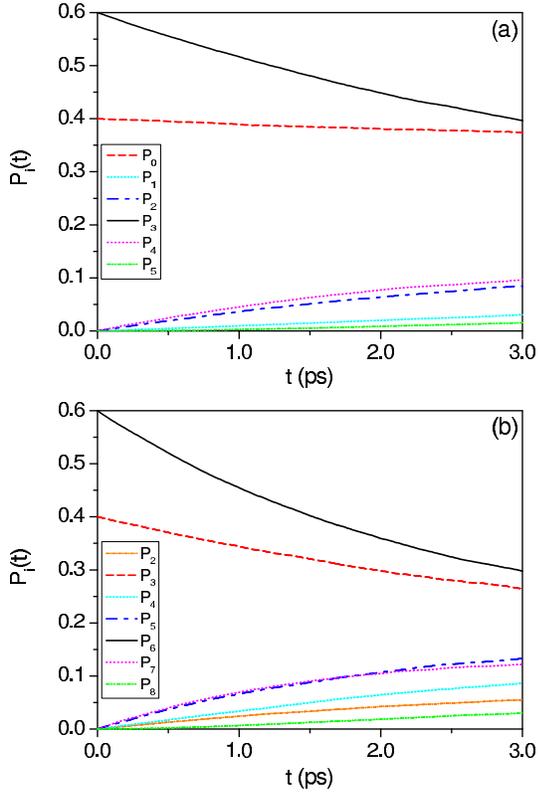}
 \caption{\label{fig6-CJC}
  Population dynamics for (a) $\Psi_{03}$ and (b) $\Psi_{36}$.
  In both graphs, the population for the higher-energy state of the
  superposition is denoted by the black solid line and the
  lower-energy state with the red dashed line; other incipient
  states have been denoted with different types of line/color
  (see the legend in each case).
  Only the populations such that $P_i(t) \ge 0.01$ at $t=3$~ps have
  been plotted.}
\end{figure}

\begin{figure}[t]
 \includegraphics[width=7cm]{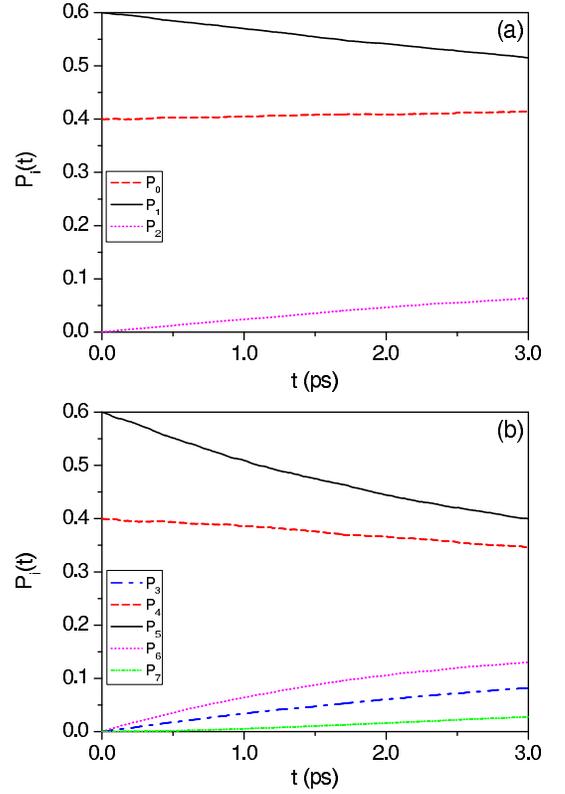}
 \caption{\label{fig7-CJC}
  Population dynamics for (a) $\Psi_{01}$ and (b) $\Psi_{45}$.
  In both graphs, the population for the higher-energy state of the
  superposition is denoted by the black solid line and the
  lower-energy state with the red dashed line; other incipient
  states have been denoted with different types of line/color
  (see the legend in each case).
  Only the populations such that $P_i(t) \ge 0.01$ at $t=3$~ps have
  been plotted.}
\end{figure}

According to eq.~\ref{trace}, $\chi$ may decay because of a change
in the populations:
\be
 P_i(t) = \rho_{S,ii}(t) = \langle\Phi_i|\hat{\rho}_{\rm S}(t)|\Phi_i\rangle ,
 \label{population}
\ee
or the suppression of the coherences:
\be
 \zeta_{ij}(t) = |\rho_{S,ij}(t)|^2
  = |\langle\Phi_i|\hat{\rho}_{\rm S}(t)|\Phi_j\rangle|^2 ,
\ee
or both effects at the same time.\cite{sanz:JCP:2006}
In order to determine which one of these possibilities occurs,
consider Fig.~\ref{fig6-CJC}, where the population dynamics
associated with the superposition states $\Psi_{03}$
(Fig.~\ref{fig6-CJC}a) and $\Psi_{36}$ (Fig.~\ref{fig6-CJC}b) are
displayed.
For simplicity, only the levels with populations equal to or larger
than 0.01 at $t=3$~ps are displayed.
We find that higher-energy levels decay faster (the decay of the
ground-level population in Fig.~\ref{fig6-CJC}a is negligible), in agreement with
the fact that purity decays faster for higher-energy superpositions.
Now, at the same time that these levels decay, others become gradually
populated.
Here, both examples confirm that the level occupation follows the
rule $m+1$, $m-1$, $n+1$, $n-1$, $m+2$, \ldots
Indeed, if two-state superpositions formed by consecutive energy levels
are considered, as seen in Fig.~\ref{fig7-CJC}, a similar rule
is found, although the level that becomes populated in the second
place corresponds to the lower-energetic state of the superposition.
This is the reason why we observe an increase of $P_0$ with time in
Fig.~\ref{fig7-CJC}b.
Nonetheless, the occupational rule of new states in this case is
$m+1$, $n-1$, $m+2$, \ldots

\begin{figure}[t]
 \includegraphics[width=7cm]{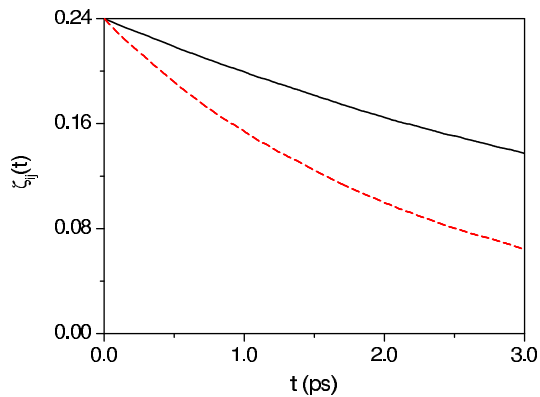}
 \caption{\label{fig8-CJC}
  Coherence dynamics, $\zeta_{ij}$, for $\Psi_{03}$ (black solid
  line) and $\Psi_{36}$ (red dashed line).
  Other incipient coherences have not been represented because they
  are negligible ($\leq 10^{-5}$).}
\end{figure}

To complete the picture, in Fig.~\ref{fig8-CJC} the coherence dynamics
for the elements $\rho_{03}$ of the (0,3)-superposition and $\rho_{36}$
for the (3,6)-superposition are displayed.
From the calculations, it was observed that only these elements
are the most strongly influenced, with their damping being again
correlated with the energy of the levels involved.
Although some other off-diagonal elements start developing, they are still
very small (smaller than $10^{-4}$) at $t=3$~ps and therefore negligible
regarding effects related to coherence dynamics.

\begin{figure}[t]
 \includegraphics[width=7cm]{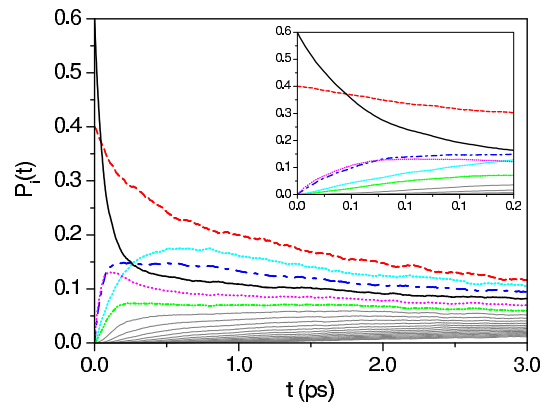}
 \caption{\label{fig9-CJC}
  Population dynamics for $\Psi_{03}$ and
  $\Lambda \approx 0.74$~\AA$^{-2}$fs$^{-1}$.
  The level populations for $\Phi_3$ and $\Phi_0$ are denoted by the
  black solid line and red dashed line, respectively.
  Other line colors/types are as in Fig.~\ref{fig6-CJC}a; gray lines
  denote populations for levels $5 \le i \le 19$, for which
  $P_i \ge 0.01$.
  Inset: detail of the short-time dynamics.}
\end{figure}

Taking these facts into account, we can conclude that the decay of
purity comes essentially from a population redistribution (diagonal
terms of the density matrix in the energy representation), including
other states that were not present in the initial superposition, and
a decay of the coherence (off-diagonal terms of the density matrix)
between the two initial states of the superposition.
As has been observed, at least in the cases analyzed (and for the
propagation time considered), although new levels start becoming
populated, no coherence appears among them.
Since the coherence between the two initial states is also gradually
lost, $\hat{\rho}_{\rm S}$ becomes asymptotically a diagonal matrix, in
agreement with the findings of the previous section (see the long-time
calculations displayed in Fig.~\ref{fig4-CJC} for large decoherence
rates).
In other words, in the long time regime, the second term (in the second
line) of eq.~\ref{trace} is expected to vanish, so that this
expression becomes
\be
 \chi_\infty \approx
  \sum_i |\langle \Phi_i|\hat{\rho}_{S,\infty}|\Phi_i\rangle|^2
  = \sum_i P_{i,\infty}^2 ,
 \label{trace2}
\ee
which is always smaller than 1 (unless only one state becomes populated
in the end).
Actually, it is also expected that population dynamics reach an
equilibrium, eventually distributing in a Boltzmann fashion.
To get a glimpse of this guess, some calculations for
$\Psi_{03}$ with $\Lambda \approx 0.74$~\AA$^{-2}$fs$^{-1}$
($5\times 10^{-3}$~a.u.) have been run in order to more quickly reach
the asymptotic regime.
These calculations seem to support the fact that populations
approach an equilibrium value, with the energy levels being
occupied in inverse relation to their energy (see Fig.~\ref{fig9-CJC}),
although at short times, the occupancy rule mentioned above is
again confirmed (see inset).


\section{Final remarks}
\label{sec5}

\vspace{-.25cm}

In this work, the quantum state diffusion approach has been used to
analyze the decoherence dynamics in the vibrational motion of I$_2$.
In spite of the limitations of this model, we have seen that it
constitutes an interesting tool to explore in a simplified manner the
decoherence dynamics of systems affected by thermal baths, without
abandoning any of the important elements involved in larger
and more detailed calculations (e.g., FB-IVR).
Indeed, the results obtained are in good agreement with
those reported in the literature from such types of
calculations.\cite{miller:JCP-1:2001,brumer-elran:JCP:2004}

Regarding the computational cost involved in the type of calculations
considered here, it is worth stressing that, compared with a standard,
wave packet unitary propagation, the multiple-realization process
involved in solving eq.~\ref{qsde} (for the same propagation time)
is essentially equivalent to performing consecutively $N$ of such unitary
propagations (the stochastic part is relatively low time-consuming,
since it does not involve additional nonlocal evaluations associated
with functions of the momentum operator $\hat{p}$, as happens with
the Hamiltonian $\hat{H}$).
This feature, together with the fact that each realization is
independent, makes the scheme suitable for parallelization, thus
increasing its time efficiency.
Compared with other methods, it is worth noticing that the
FB-IVR calculations used in Wang et al.\cite{miller:JCP-1:2001}
required a total of $5\times 10^4$ to $10^5$ $2(M+1)$-dimensional
realizations in order to reach convergence.
Each one of these realizations involves the two classical degrees of
freedom of the system ($x,p$) and those of the oscillators ($Q_i,P_i$),
with $i=1, 2 \ldots , M$ (about 20-40 oscillators were needed by the
authors to properly describe the continuous spectral density assumed
in Wang et al.\cite{miller:JCP-1:2001}).
Putting aside the time consumed in evaluating the monodromy matrix
elements required by the FB-IVR, the computation of such a number of
$2(M+1)$-dimensional classical trajectories is relatively
demanding,\cite{brumer-elran:JCP:2004} even though the oscillators are
not coupled among themselves, as happens in more realistic bath
models.\cite{sanz:PRE:2012,brumer-elran:JCP:2013}
In this regard, solving eq.~\ref{qsde} is advantageous both
computationally and interpretively, since it is less time-consuming
and provides a similar degree of accuracy (when properly tuned),
as seen in Section~\ref{sec41}.

There are a few issues that are left open in this work but on which
there is currently some preliminary work in progress.
First, the application of this methodology to other contexts of
interest, for example, scattering systems, such as slit
systems\cite{miller:JCP-2:2001,hornberger:PRL:2003} and
atom-surface collisions,\cite{pollak:JCP:2009,pollak:PRL:2010} where
it is shown that the standard textbook guess of only varying some
typical quantum parameter to reach the classical limit is not
valid.\cite{sanz:EPL:2001}
Second, and more importantly, how to link in a systematic fashion this
approach to more exact calculations and to more realistic
systems.\cite{sanz:PRE:2012,brumer-elran:JCP:2013}
In this work, the value of the decoherence rate $\Lambda$ has been
changed because we had the freedom to choose the coefficient
$\xi = \eta T$ and compare it with results available in the literature.
However, it is highly desirable to find out a way to determine this
quantity in a unique manner once the nature of the bath is specified
(e.g., type of system--bath and bath--bath interactions) and its physical
conditions are defined (e.g., temperature).
Notice that depending on the bath nature, the influence over the system
will be different\cite{sanz:PRE:2012} even for the same (bath) physical
conditions.
In this sense, the form of the Lindblad operators should also be
analyzed, since the position operator may be valid for bilinear
couplings (as is the case of the Caldeira--Leggett model), but not
in more complex situations (e.g., interactions of diatomics with rare
gas liquids or solid matrices).


\vspace{-.5cm}

\acknowledgments

\vspace{-.25cm}

This work is dedicated to Paul Brumer, a good man and an excellent
teacher, with all my affection and admiration.
The author thanks the Chemical Physics Theory Group of the University
of Toronto for kind access to its computational facilities, as well as
the Ministerio de Econom\'{\i}a y Competitividad (Spain) for economical
support under Project FIS2011-29596-C02-01 and a ``Ram\'on y Cajal''
Research Grant.



\begin{thebibliography}{46}
\bibitem{footnote}
To be more precise in this regard, rather than using the term
``classical'', it should be said in analogy to the solutions of ordinary
differential equations (of which the equations of classical mechanics
are just a particular type).
Partial differential equations, like Schr\"odinger's equation or
eq.~\ref{itostoch}, describe the propagation of field-type variables
and therefore the concept of trajectory is ambiguous unless some
criterion is properly defined.
For example, within quantum state diffusion, trajectories are defined
as the expectation value of some operator.
This means that, given a particular initial wave function, there is
always a multiplicity of trajectories, as many as realizations are
considered.
If $\hat{\mathcal{O}} = \hat{x}$, each trajectory corresponds to the
position expectation value obtained from each independent, stochastic
realization of the state vector $|\Psi\rangle_i$, which here is
denoted by $\langle \hat{x}\rangle_i(t)$ (see Section~\ref{sec41}).
As another example, in Bohmian mechanics \cite{sanz-bk-1,sanz-bk-2} for
the same initial wave function one has that for a single realization
there are as many trajectories as chosen initial conditions;
these initial conditions are taken in such a way that the positions are
distributed according to the corresponding probability density and the
momenta are fixed by the initial phase of the wave function.
In this case, trajectories would be labeled as $x_i(t)$ (like
``classical'' trajectories), with the position expectation value being
the average over a large ensemble of these trajectories, i.e.,
$\langle \hat{x}\rangle(t) = (1/N)\sum_{i=1}^N x_i(t)$.
\end{thebibliography}


\end{document}